\documentclass[]{elsart}
\usepackage[dvips]{graphicx}
\begin{document}
\begin{frontmatter}
%\twocolumn[\hsize\textwidth\columnwidth\hsize\csname @twocolumnfalse\endcsname
\title{A mechanism for randomness}  
\author[ivic,trieste]{J. A. Gonz\'alez},
\author[usb,trieste]{L. I. Reyes},
\author[ivic,usb]{J. J. Su\'{a}rez},
\author[usb]{L. E. Guerrero}%
, and
\author[usb]{G. Guti\'{e}rrez}
\address[ivic]{Centro de F\'{\i}%
sica, Instituto Venezolano de Investigaciones
Cient\'{\i}%
ficas, Apartado Postal 21827, Caracas 1020-A, Venezuela}%

\address[trieste]{The Abdus Salam International Centre for Theoretical Physics,
Strada Costiera 11, 34100, Trieste, Italy}%

\address[usb]{Departamento de F\'{\i}sica, Universidad Sim\'on Bol\'{\i}%
var, Apartado Postal 89000, Caracas 1080-A, Venezuela}%

%\thanks[add]{Corresponding author. Fax: +58-212-5041148; e-mail: jorge@pion.ivic.ve}%

\begin{abstract}%

We investigate explicit functions that can produce truly random numbers. We use 
the analytical properties of the explicit functions to show that certain class
of autonomous dynamical systems can generate random dynamics. This dynamics 
presents fundamental differences with the known chaotic systems. We present real
physical systems that can produce this kind of random time-series. We report the
results of real experiments with nonlinear circuits containing direct evidence
for this new phenomenon. In particular, we show that a Josephson junction
coupled to a chaotic circuit can generate unpredictable dynamics. Some 
applications are discussed.
\end{abstract}%

\end{frontmatter}

\section{Introduction}

Over the last three decades or so, a revolution has happened in the
development of science. We are talking about Chaos theory
\cite{Lorenz,Li,Feigenbaum,May,Ford,Crutchfield,Grebogi,Ott,Takens,Ruelle}. In
the chaotic regime, the behavior of a deterministic system appears random.
This finding has forced many experimentalists to re-examine their data to
determine whether some of the random behaviors attributed to noise are due to
deterministic chaos instead.

Chaos theory has been successfully applied to many scientific and practical
situations
\cite{Lorenz,Li,Feigenbaum,May,Ford,Crutchfield,Grebogi,Ott,Takens,Ruelle}.

In the philosophical realm, however, the importance of this development was
that chaos theory seemed to offer scientists the hope that almost ``any''
random behavior observed in Nature could be described using low-dimensional
chaotic systems. Random-looking information gathered in the past (and shelved
because it was assumed to be too complicated) perhaps could now be explained
in terms of simple laws.

The known chaotic systems are not random \cite{Farmer}. If the previous values
of a time-series determine the future values, then even if the dynamical
behavior is chaotic, the future may, to some extent, be predicted from the
behavior of past values that are similar to those of the present. The
so-called ``unpredictability'' in the known chaotic systems is the result of
the sensitive dependence on initial conditions. It is not an absolute unpredictability.

Truly random systems are different from the chaotic ones. Past sequences of
values of a random dynamical variable that are similar to present ones tell as
much or little about the next value as about the next hundredth value. The
so-called nonlinear forecasting methods for distinguishing chaos from random
time-series are based on these ideas \cite{Farmer}.

Recently, we have introduced explicit functions that produce truly random
sequences \cite{Gonzalez1,Nazareno,Gonzalez2}. For instance, let us define the function:%

\begin{equation}
X_{n}=\sin^{2}\left(  \theta\pi z^{n}\right)  ,\label{exp1}%
\end{equation}
where $z$ is a real number and $\theta$ is a parameter.

For an integer $z>1,$ this is the solution to some chaotic maps
\cite{Gonzalez1,Nazareno,Gonzalez2} (see Fig. 1(a)). For a non-integer $z$,
function (\ref{exp1}) can produce truly unpredictable random sequences whose
values are independent.

\begin{figure}[ptb]
%%\epsfysize=12.0truecm
%%\epsfxsize=6.0truecm
%%\rotatebox{270}{\epsffile{fig1.eps}}
\centerline{\includegraphics[width=6.0cm,angle=-90]{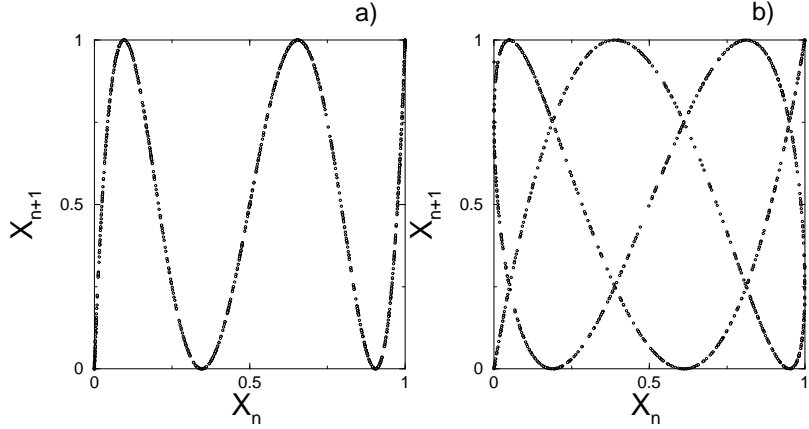}}  \vspace
{5mm}\caption{First-return maps produced by function (\ref{exp1}): (a) $z=5$;
(b) $z=7/3$.}%
\label{Fig.1}%
\end{figure}

Functions (\ref{exp1}) with non-integer $z$ cannot be expressed as a map of type%

\begin{equation}
X_{n+1}=f\left(  X_{n},X_{n-1},...,X_{n-r+1}\right)  .\label{exp2}%
\end{equation}

In the present letter we address the following question: can an autonomous
dynamical system with several variables produce a random dynamics similar to
that of function (\ref{exp1})? We will present several dynamical systems with
this kind of behavior. We will report the results of real experiments with
nonlinear circuits, which contain direct evidence for this new phenomenon. We
discuss some applications.

\section{Random functions}

Let us discuss first some properties of function (\ref{exp1}). We will present
here a short proof of the fact that the sequences generated by functions
(\ref{exp1}) are unpredictable from the previous values. This proof is
presented here for the first time. However, a more detailed discussion of the
properties of these functions (including statistical tests) can be found in
Ref. \cite{Gonzalez1,Nazareno,Gonzalez2}.

Let $z$ be a rational number expressed as $z=p/q$, where $p$ and $q$ are
relative prime numbers.

We are going to show that if we have $m+1$ numbers generated by function
(\ref{exp1}): $X_{0},X_{1},X_{2},X_{3},...,X_{m}$ ($m$ can be as large as we
wish), then the next value $X_{m+1}$ is still unpredictable. This is valid for
any string of $m+1$ numbers.

Let us define the following family of sequences:%

\begin{equation}
X_{n}^{(k,m)}=\sin^{2}\left[  \pi\left(  \theta_{0}+q^{m}k\right)  \left(
\frac{p}{q}\right)  ^{n}\right]  ,\label{exp3}%
\end{equation}
where $k$ is an integer. The parameter $k$ distinguishes the different sequences.

For all sequences parametrized by $k$, the first $m+1$ values are the same.
This is so because%

\begin{equation}
X_{n}^{(k,m)}=\sin^{2}\left[  \pi\theta_{0}\left(  \frac{p}{q}\right)
^{n}+\pi kp^{n}q^{(m-n)}\right]  =\sin^{2}\left[  \pi\theta_{0}\left(
\frac{p}{q}\right)  ^{n}\right]  ,\label{exp4}%
\end{equation}
for all $n\leq m$. Note that the number $kp^{n}q^{(m-n)}$ is an integer for
$n\leq m$. So we can have infinite sequences with the same first $m+1$ values.

Nevertheless, the next value%

\begin{equation}
X_{m+1}^{(k,m)}=\sin^{2}\left[  \pi\theta_{0}\left(  \frac{p}{q}\right)
^{m+1}+\frac{\pi kp^{m+1}}{q}\right] \label{exp5}%
\end{equation}
is uncertain.

In general, $X_{m+1}^{(k,m)}$ can take $q$ different values. These $q$ values
can be as different as $0,1/2,\sqrt{2}/2,1/e,1/\pi,$ or $1$. From the
observation of the previous values $X_{0},X_{1},X_{2},X_{3},...,X_{m}$, there
is no method for determining the next value.

This result shows that for a given set of initial conditions, there exists
always an infinite number of values of $\theta$ that satisfy those initial
conditions. The time-series produced for different values of $\theta$
satisfying the initial conditions is different in most of the cases. Even if
the initial conditions are exactly the same, the following values are
completely different. This property is, in part, related to the fact that the
equation $\sin^{2}\theta=\alpha$, where $0\leq\alpha\leq1$, possesses infinite
solutions for $\theta$.

We should stress that from the observation of a string of values $X_{0}%
,X_{1},X_{2},X_{3},...,X_{m}$ generated by function (\ref{exp1}) it is
impossible to determine which value of $\theta$ was used.

Figures 1(a) and 1(b) show the first-return maps for $z=5$ and $z=7/3$.

For $z$ irrational (we exclude the numbers of type $z=m^{1/k}$), the numbers
generated by function (\ref{exp1}) are completely independent (see Fig. 2(a)
that shows the first-return map for $z=e$). After any string of $m+1$ numbers
$X_{0},X_{1},X_{2},X_{3},...,X_{m}$, the next outcome $X_{m+1}$ can take
infinite different values.

\begin{figure}[ptb]
%%\epsfysize=12.0truecm
%%\epsfxsize=6.0truecm
%%\rotatebox{270}{\epsffile{fig2.eps}}
\centerline{\includegraphics[width=6.0cm,angle=-90]{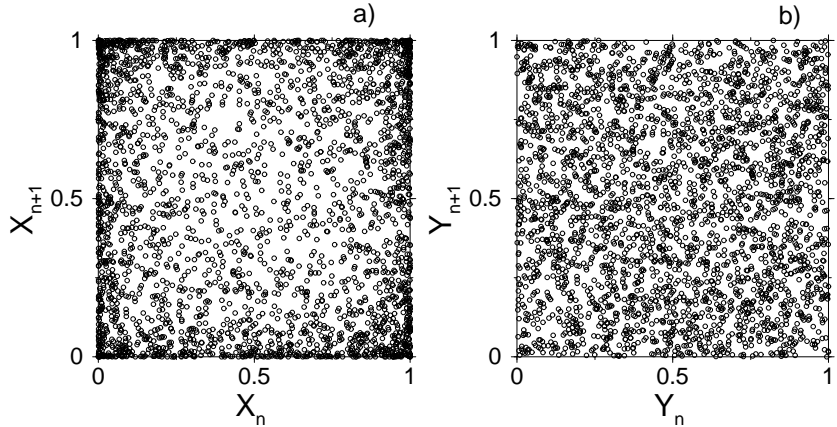}}  \vspace
{5mm}\caption{First-return maps produced by function (\ref{exp1}): (a) $z=e$
(irrational); (b) $z=e$, $Y_{n}=\left(  2/\pi\right)  \arcsin\left(
X_{n}^{1/2}\right) $.}%
\label{Fig.2}%
\end{figure}

The numbers produced by function (\ref{exp1}) are random but are not
distributed uniformly. The probability density behaves as $P(X)\sim
1/\sqrt{X(1-X)}$. If we need uniformly distributed random numbers, we should
make the following transformation $Y_{n}=\frac{2}{\pi}\arcsin\sqrt{X_{n}}$. In
this case $P(Y)=const$ (see Fig. 2(b)).

It is important to mention here that the argument of function (\ref{exp1})
does not need to be exponential all the time, for $n\rightarrow\infty$. In
fact, a set of finite sequences (where each element-sequence is unpredictable,
and the law for producing a new element-sequence cannot be obtained from the
observations) can form an infinite unpredictable sequence. See the discussion
in the following paragraph.

So if we wish to produce random sequences of very long length, we can
determine a new value of parameter $\theta$ after a finite number $N$ of
values of $X_{n}$. This procedure can be repeated the desired number of times.
It is important to have a nonperiodic method for generating the new value of
$\theta$. For example, we can use the following method in order to change the
parameter $\theta$ after each set of $N$ sequence values. Let us define
$\theta_{s}=AW_{s}$, where $W_{s}$ is produced by a chaotic map of the form
$W_{s+1}=f(W_{s})$; $s$ is the order number of $\theta$ in a way that $s=1$
corresponds to the $\theta$ used for the first set of $N$ values of $X_{n}$,
$s=2$ for the second set, etc. The inequality $A>1$ should hold to ensure the
absolute unpredictability. In this case, from the observation of the values
$X_{n}$, it is impossible to determine the real value of $\theta$.

After a careful analysis of functions (\ref{exp1}), we arrive at the
preliminary conclusion that (to produce unpredictable dynamics) the main
characteristics for any functions are the following: the function should be
able to be re-written in the form%

\begin{equation}
X_{n}=h\left(  f\left(  n\right)  \right)  ,\label{exp6}%
\end{equation}
where the argument function $f\left(  n\right)  $ grows exponentially and the
function $h(y)$ should be finite and periodic. This result allows us to
generalize this behavior to other functions as the following%

\begin{equation}
X_{n}=P\left(  \theta z^{n}\right)  ,\label{exp7}%
\end{equation}
where $P(t)$ is a periodic function.

However, a more deep analysis shows that (to produce complex behavior) the
function $f(n)$ does not have to be exponential all the time, and function
$h(y)$ does not have to be periodic. In fact, it is sufficient for function
$f(n)$ to be a finite nonperiodic oscillating function which possesses
repeating intervals with finite exponential behavior. For instance, this can
be a chaotic function. On the other hand, function $h(y)$ should be
noninvertible. In other words, it should have different maxima and minima in
such a way that equation $h(y)=\alpha$ (for some specific interval of $\alpha
$, $\alpha_{1}<\alpha<\alpha_{2}$) possesses several solutions for $y$.

\section{Autonomous dynamical systems}

The following autonomous dynamical system can produce truly random dynamics:
\begin{equation}
X_{n+1} = \cases{aX_n, &if $X_n<Q$,\cr bY_n, &if $X_n>Q$,} \label{exp8}%
\end{equation}
\begin{equation}
Y_{n+1}=cZ_{n},\label{exp9}%
\end{equation}
\begin{equation}
Z_{n+1}=\sin^{2}\left(  \pi X_{n}\right) ,\label{exp10}%
\end{equation}

Here $a>1$ can be an irrational number, $b>1$, $c>1$. We can note that for
$0<X_{n}<Q$, the behavior of function $Z_{n}$ is exactly like that of function
(\ref{exp1}).

For $X_{n}>Q$ the dynamics is reinjected to the region $0<X_{n}<Q$ with a new
initial condition. While $X_{n}$ is in the interval $0<X_{n}<Q$, the dynamics
of $Z_{n}$ is unpredictable as it is function (\ref{exp1}). Thus, the process
of producing a new initial condition through the equation (\ref{exp9}) is random.

If the only observable is $Z_{n}$, then it is impossible to predict the next
values of this sequence using only the knowledge of the past values.

An example of the dynamics produced by the dynamical system (\ref{exp8}%
)-(\ref{exp10}) is shown in Fig. 3. If we apply the nonlinear forecasting
method analysis to a common chaotic system, then the prediction error
increases with the number of time-steps into the future. On the other hand,
when we apply this method to the time-series produced by system (\ref{exp8}%
)-(\ref{exp10}), the prediction error is independent of the time-steps into
the future, as in the case of a random sequence. Other very strong
methods\cite{Grassberger}, which allow to distinguish between chaos and random
noise, produce the same result.

\begin{figure}[ptb]
%%\epsfysize=12.0truecm
%%\epsfxsize=6.0truecm
%%\rotatebox{270}{\epsffile{fig3.eps}}
\centerline{\includegraphics[width=6.0cm,angle=-90]{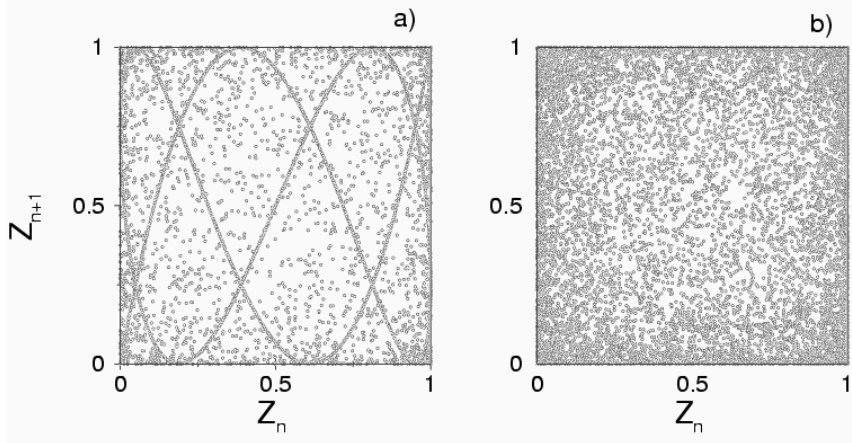}}  \vspace
{5mm}\caption{First-return maps produced by the dynamics of the dynamical
system (\ref{exp8})-(\ref{exp10}). (a) $a=7/3$, $b=171$, $c=1.5$, $Q=1000/a$;
(b) $a=e$ (irrational), $b=171$, $c=1.5$, $Q=1000/a$.}%
\label{Fig.3}%
\end{figure}

Here we should make an important remark. Mathematical models can be of
different types. For instance, natural phenomena can be described by
differential equations, difference equations, cellular automata, neural
networks, etc.

Even explicit functions can be mathematical models. If we consider function
(\ref{exp1}) as the mathematical model of some dynamical process, then this
process can be completely random. If all we know is the series of outcomes
$X_{0},X_{1},X_{2},...$, then it does not matter how many values we already
have, the next value is unpredictable. However, \ this is because from the
observation of the values $X_{0},X_{1},X_{2},...,X_{m}$, it is impossible to
determine the ``variable'' $Y_{n}=\theta\pi z^{n}$. This could be considered
as a ``hidden variable''. Of course, we cannot say that this is just a problem
of hidden variables, because not for any hidden variable $Y_{n}$, the function
$X_{n}=\sin^{2}\left(  Y_{n}\right)  $ is a random system. Here we have
obtained necessary conditions for this phenomenon to occur.

Let us extend this analysis to the dynamical system (\ref{exp8})-(\ref{exp10}%
). In this case the completely random variable is $Z_{n}$. The role of
``hidden variable'' is played by $X_{n}$. If one could observe the series
$X_{n}$, then the complete randomness of the data set is lost. We could say
that some projection of the complete set of variables onto a proper subset is
necessary for the effect.

It is important to notice that the dynamical system (\ref{exp8})-(\ref{exp10})
is a well-posed set of difference equations with unique forward time
evolution. Given the initial conditions ($X_{0}$, $Y_{0}$, $Z_{0}$), the
future evolution of the dynamics is fully determined. However, if the only
observable is $Z_{n}$, then this variable will behave as a completely random
time-series. How can one reconcile the unique forward time evolution with this
randomness? The answer is related to the fact that only a subset of the
variables are observed.

With this result we are uncovering a new mechanism for generating random
dynamics. This is a fundamental result because it is very important to
understand different mechanisms by which the natural systems can produce truly
random (not only chaotic) dynamics.

\section{Pseudorandom number generators}

There is a large literature dedicated to pseudorandom number generators (see
e.g.
\cite{Ferrenberg,James1,James2,Collins,Knuth,Park,Doob,Mark,D'Souza,Nogues,Fisher,Marsaglia1,Marsaglia2,Marsaglia3,Vattulainen,Shchur,Li2,Phatak,Kautz,Proykova,Stojanovski1,Stojanovski2,Bernardini}
and references quoted therein). A very fine theory has been developed in this
area and this theory has produced many important results.

However, we should say here that the known pseudorandom number generators are
not supposed to generate truly random numbers.

In his important review article \cite{James1}, James says : ``Truly random
numbers are unpredictable in advance and must be produced by a random physical
process, such as radioactive decay''.

In fact, pseudorandom numbers are produced using recurrence relations, and are
therefore not truly random \cite{James1,James2,D'Souza,Nogues,Fisher,Phatak}.

D'Souza et al \cite{D'Souza} say in their paper: ``Pseudorandom number
generators are at best a practical substitute, and should be generally tested
for the absence of undesired correlations''.

Many known pseudorandom number generators are based on maps of type:%

\begin{equation}
X_{n+1}=f\left(  X_{n},X_{n-1}...,X_{n-r+1}\right)  .\label{exp19}%
\end{equation}

Now we will present the maps behind some of the most famous and best
pseudorandom number generators.

\textit{Multiplicative linear congruential generators}
\cite{Ferrenberg,James1} are defined by the following equation:%

\begin{equation}
X_{n+1}=\left( aX_{n}+c\right)  \operatorname{mod}m\label{exp20}%
\end{equation}

Some famous values for these parameters are the following: $a=23$,
$m=10^{8}+1$, $c=0$; $a=65539$, $m=2^{29}$, $c=0$; $a=69069$, $m=2^{32}$,
$c=1$; $a=16807$, $m=2^{31}-1$, $c=0$; $a=1664525$, $m=2^{32}$, $c=0$ (this is
the best generator for $m=2^{32}$, according to the criteria of Knuth
\cite{Knuth}).

The \textit{Fibonacci-like generators} obey the following equation:%

\begin{equation}
X_{n+1}=\left(  X_{n-p}\odot X_{n-q}\right)  \operatorname{mod}%
m,\label{exp21}%
\end{equation}
where $\odot$ is some binary or logical operation. For instance, $\odot$ can
be addition, subtraction or exclusive-or.

Other \textit{extended algorithms} use equations as the following:%

\begin{equation}
X_{n+1}=\left(  aX_{n}+bX_{n-1}+c\right)  \operatorname{mod}m.\label{exp22}%
\end{equation}

The \textit{add-and-carry generators} are defined as:%

\begin{equation}
X_{n+1}=\left(  X_{n-r}\pm X_{n-s}\pm c\right)  \operatorname{mod}%
m.\label{exp23}%
\end{equation}

Among the high quality generators investigated in the famous paper
\cite{Ferrenberg} are the following%

\begin{equation}
X_{n+1}=\left(  16807X_{n}\right)  \operatorname{mod}\left(  2^{31}-1\right)
,\label{exp24}%
\end{equation}%

\begin{equation}
X_{n+1}=\left(  X_{n-103}.XOR.X_{n-250}\right)  ,\label{exp25}%
\end{equation}%

\begin{equation}
X_{n+1}=\left(  X_{n-1063}.XOR.X_{n-1279}\right)  ,\label{exp26}%
\end{equation}
where $.XOR.$ is the bitwise exclusive OR operator,%

\begin{equation}
X_{n}=\left(  X_{n-22}-X_{n-43}-c\right)  ,\label{exp27}%
\end{equation}
here for $X_{n}\geq0$, $c=0$, and for $X_{n}<0$, $X_{n}=X_{n}+\left(
2^{32}-5\right)  $, $c=1$.

All known generators (in some specific physical calculations) give rise to
incorrect results because they deviate from randomness
\cite{Ferrenberg,D'Souza,Nogues}.

The problem is that these algorithms are predictable.

An example of this can be found in the work of Ferrenberg et al
\cite{Ferrenberg}. They found that high quality pseudorandom number generators
can yield incorrect answers due to subtle correlations between the generated numbers.

Suppose we have an ideal generator for truly random numbers. In this case, no
matter how many numbers we have generated, the value of the next number will
be still unknown. That is, there is no way to write down a formula that will
give the value of the next number in terms of the previous numbers, no matter
how many numbers have been already generated.

The authors of paper \cite{Ferrenberg} related the errors in the simulations
to the dependence in the generated numbers. Indeed, they are all based on maps
of type (\ref{exp19}).

In the present paper we have shown that both the sequence of numbers $X_{n}$
defined by function (\ref{exp1}) and the sequence of numbers $Z_{n}$ defined
by the dynamical system (\ref{exp8})-(\ref{exp10}) cannot be expressed as a
map of type (\ref{exp19}). In fact, these numbers are unpredictable and the
next value cannot be determined as a function of the previous values.

Recently, simulations of different physical systems have become the
``strongest'' tests for pseudorandom number generators. Among these systems
are the following: the two-dimensional Ising model \cite{Ferrenberg},
ballistic deposition \cite{D'Souza}, and random walks \cite{Nogues}.

Nogu\'{e}s et al \cite{Nogues} have found that using common pseudorandom
number generators, the produced random walks present symmetries, meaning that
the generated numbers are not independent.

\begin{figure}[ptb]
%%\epsfysize=12.0truecm
%%\epsfxsize=12.0truecm
%%\epsffile{fig4.eps}
\centerline{\includegraphics[width=8.0cm,angle=0]{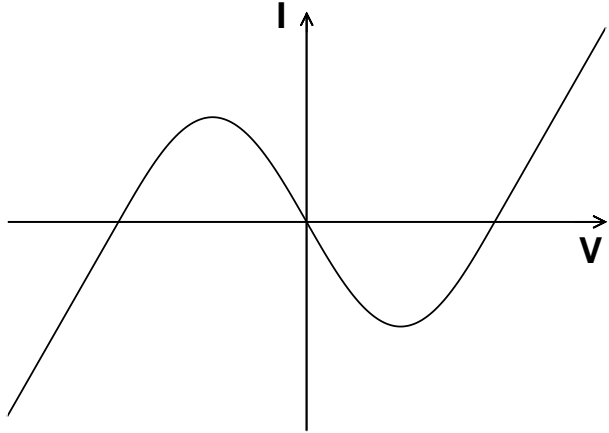}}  \vspace
{5mm}\caption{Noninvertible I-V characteristic. Two extrema.}%
\label{Fig.4}%
\end{figure}

On the other hand, the logarithmic plot of the mean distance versus the number
of steps $N$ is not a straight line (as expected theoretically) after
$N>10^{5}$ (in fact, it is a rapidly decaying function).

D'Souza et al \cite{D'Souza} use ballistic deposition to test the randomness
of pseudorandom number generators. They found correlations in the pseudorandom
numbers and strong coupling between the model and the generators (even
generators that pass extensive statistical tests).

One consequence of the Kardar-Parisi-Zhang theory is that the steady state
behavior for the interface fluctuations (in ballistic deposition in one
dimension) should resemble a random walk. Thus, a random walk again serves as
a good test for pseudorandom numbers.

We have produced random walks using the numbers generated by our systems. The
produced random walks possess the correct properties, including the mean
distance behavior $\langle d^{2}\rangle\sim N$.

The present paper is not about random number generators. In the present paper
we discuss a new phenomenon: the fact that unperturbed physical systems can
produce truly random dynamics.

Of course, one of the applications of this phenomenon is random number generation.

The art of random number generation requires more than the randomness of the
generated numbers. It requires good programming skills and techniques to
obtain the desired distributions for the numbers.

The functions and systems described in this paper can be used to create very
good random number generators. Algorithms designed for this purpose along with
the statistical tests will be published elsewhere.

In this section we only wished to present a theoretical comparison between the
sequences produced by the pseudorandom-number-generator algorithms
(\ref{exp19})-(\ref{exp27}) and the systems described in the present paper.

\section{Experiments}

When the input is a normal chaotic signal and the system is an electronic
circuit with the I-V characteristics shown in Figures 4 and 5, then the output
will be a very complex signal.

%~\vspace{.3cm}
\begin{figure}[ptb]
%%\epsfysize=6.0truecm
%%\epsfxsize=6.0truecm
%%\rotatebox{270}{\epsffile{fig5.eps}}
\centerline{\includegraphics[width=8.0cm,angle=0]{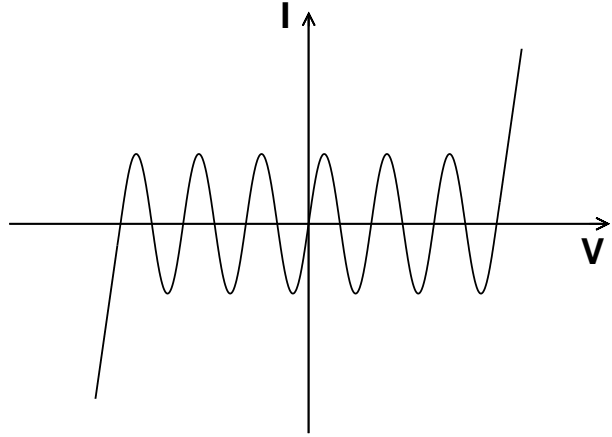}}  \vspace
{5mm}\caption{Noninvertible I-V characteristic. Many extrema.}%
\label{Fig.5}%
\end{figure}

In Ref. \cite{Chua} a theory of nonlinear circuits is presented. There we can
find different methods to construct circuits with these I-V characteristic curves.

The scheme of this composed system is shown in Fig. 6. A set of equations
describing this dynamical system is the following:%

\begin{equation}
X_{n+1}=F_{1}\left(  X_{n},Y_{n}\right)  ,\label{exp11}%
\end{equation}%

\begin{equation}
Y_{n+1}=F_{2}\left(  X_{n},Y_{n}\right)  ,\label{exp12}%
\end{equation}%

\begin{equation}
Z_{n+1}=g\left(  X_{n}\right)  ,\label{exp13}%
\end{equation}
where the equations (\ref{exp11}) and (\ref{exp12}) describe a normal chaotic
dynamics where the variable $X_{n}$ presents intermittent intervals with a
truncated exponential behavior and $g\left(  X_{n}\right)  $ is a function
with several maxima and minima as that shown in Fig. 5.

\begin{figure}[ptb]
%%\epsfysize=24.0truecm
%%\epsfxsize=12.0truecm
%%\epsffile{fig6.eps}
\centerline{\includegraphics[width=12.0cm,angle=0]{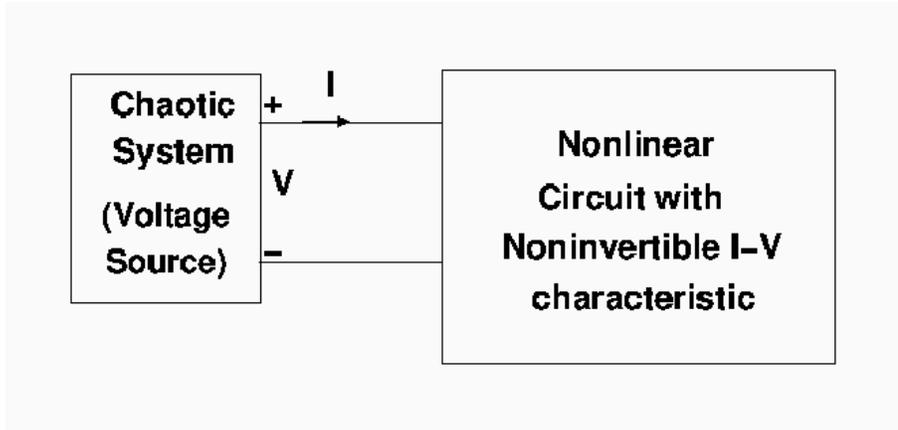}}  \vspace
{5mm}\caption{Scheme of a nonlinear system where a chaotic voltage source is
used as the input signal for a nonlinear circuit with a noninvertible I-V
characteristic.}%
\label{Fig.6}%
\end{figure}

Figures 7(a) and 7(b) show nonlinear circuits that can be used as the
nonlinear system shown on the right of the scheme of Fig. 6.

\begin{figure}[ptb]
%%\epsfysize=24.0truecm
%%\epsfxsize=12.0truecm
%%\epsffile{fig7.eps}
\centerline{\includegraphics[width=12.0cm,angle=0]{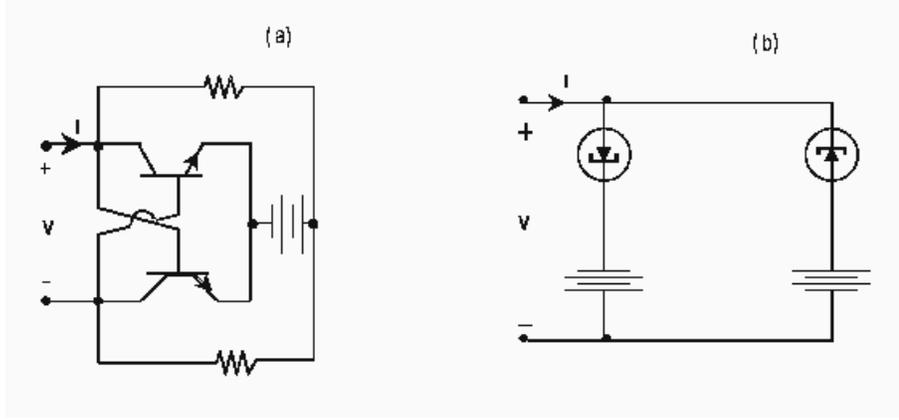}}  \vspace
{5mm}\caption{Nonlinear circuits with noninvertible I-V characteristics. (a)
The resistors possess $R=2.2k\Omega$, the source voltage in the battery is
$10V$ and the twin transistors are $2N2222$ with $\beta=140$). The I-V
characteristic of this circuit is shown in Fig. 4. (b) Another circuit with a
similar I-V characteristic.}%
\label{Fig.7}%
\end{figure}

The system on the left of the scheme can be a chaotic circuit, e.g. the Chua's
circuit \cite{Matsumoto}.

We have constructed a circuit similar to the one shown in Fig. 7(a). We
produced chaotic time-series using a common nonlinear map and then we
transformed them into analog signals using a converter. This analog signals
were introduced as the voltage-input to the circuit shown in Fig. 7(a).
Similar results are obtained when we take the input signal from a chaotic
electronic circuit.

The set of equations that describes one of our experimental situations is the following:%

\begin{equation}
X_{n+1}=aX_{n}\left[  1-\Theta(X_{n}-q)\right]  +bY_{n}\Theta(X_{n}%
-q),\label{exp14}%
\end{equation}%

\begin{equation}
Y_{n+1}=\sin^{2}\left[  d\arcsin\sqrt{Y_{n}}\right]  ,\label{exp15}%
\end{equation}%

\begin{equation}
Z_{n+1}=4W_{n}^{3}-3W_{n},\label{exp16}%
\end{equation}
where $W_{n}=\frac{2X_{n}}{s}-1$, $q=\frac{s}{a}$, $s=10$, $b=7$,
$a=\frac{\pi}{2}$, $d=3$, $\Theta(x)$ is the Heaviside function.

The first-return maps of the sequence $Z_{n}$ produced by the theoretical
model (\ref{exp14})-(\ref{exp16}) and the experimental time-series produced by
the nonlinear system of Figs. 6 and 7(a) are shown in Figures 8(a) and 8(b).

\begin{figure}[ptb]
%%\epsfysize=12.0truecm
%%\epsfxsize=6.0truecm
%%\rotatebox{270}{\epsffile{fig8.eps}}
\centerline{\includegraphics[width=6.0cm,angle=-90]{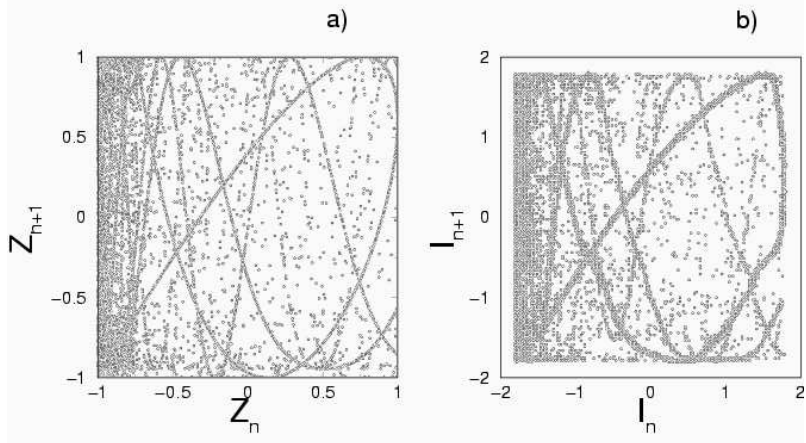}}  \vspace
{5mm}\caption{Modelling vs experiment. (a) Numerical simulation of the
dynamical system (\ref{exp14})-(\ref{exp16}); (b) first-return map produced
with the real data (current measurements) from experiments using the scheme of
Fig. 6, where the circuit of the right is the one of Fig. 7(a).}%
\label{Fig.8}%
\end{figure}

When the nonlinear circuit has an I-V characteristic with many more maxima and
minima, e.g. Fig. 5 (and this can be done in practice, see Ref. \cite{Chua}),
we can produce a much more complex dynamics.

Nonlinear chaotic circuits can be described successfully by discrete maps as
the equations (\ref{exp11})-(\ref{exp13}) (see e.g. \cite{Rodriguez}).

However, in some cases it can be very helpful to have a physical situation
with a model based on a set of well-posed ordinary differential equations.

In Ref. \cite{Huang} we can find several models for chaotic circuits as the following:%

\begin{equation}
\frac{dX}{dt}=\alpha\left[  f\left(  x\right)  -Y\right]  ,\label{exp32}%
\end{equation}%

\begin{equation}
\frac{dY}{dt}=Y-X-Z,\label{exp33}%
\end{equation}%

\begin{equation}
\frac{dZ}{dt}=\beta Y+\gamma Z,\label{exp34}%
\end{equation}%

\begin{equation}
\frac{dW}{dt}=g\left(  X\right)  .\label{exp35}%
\end{equation}
where $f\left(  x\right)  =-X^{3}+cX$, $g\left(  X\right)  =\sum
\limits_{i=1}^{N}a_{i}x^{i}$.

A comparison of different time-series is shown in Fig. 9. The fixed parameter
values are: $\alpha=285.714$, $\beta=1499.25$, $c=0.144$, $\gamma=-0.51325$.
Note that when the I-V characteristic of the circuit shown on the right of
Fig. 6 is a function with many extrema, the produced time-series is more
complex (see Fig. 9(b)).

Using our theoretical results we can make a very important prediction here. A
nonlinear physical system constructed with chaotic circuits and a Josephson
junction \cite{Barone} can be an ideal experimental setup for the random
dynamics that we are presenting here.

It is well-known that the current in a Josephson junction may be written as%

\begin{equation}
I=I_{c}\sin\phi,\label{exp 17}%
\end{equation}
where%

\begin{equation}
\frac{d\phi}{dt}=kV\label{exp 18}%
\end{equation}
Here $\phi$ is the phase and $V$ is the voltage across the junction. Note that
Nature has provided us with a phenomenon where the sine-function is intrinsic.
Although we have already explained that other noninvertible functions can
produce similar results, it is remarkable that we can use this very important
physical system to investigate the real consequences of our results with
function (\ref{exp1}). In a superconducting Josephson junction $k$ is defined
through the fundamental constants $k=2e/\hbar$. However, in the last decades
there have been a wealth of experimental work dedicated to the creation \ of
electronic analogs that can simulate the Josephson junction
\cite{Werthamer,Hamilton,Bak,Magerlein}. In that case $k$ can be a parameter
with different numerical values.

We have performed real experiments with a nonlinear chaotic circuit coupled to
an analog Josephson junction.

In our experiments we have used the Josephson junction analog constructed by
Magerlein \cite{Magerlein}. This is a very accurate device that has been found
very useful in many experiments for studying junction behavior in different
circuits. The junction voltage is integrated using appropriate resetting
circuitry to calculate the phase $\phi$, and a current proportional to
$\sin\phi$ can be generated. The circuit diagram can be found in
\cite{Magerlein}.

The parameter $k$ is related to certain integrator time constant $RC$ in the
circuit. So we can change its value. This is important for our experiments. We
need large values of $k$ in order to increase the effective domain of the sine
function. In other words, we need the argument of the sine function to take
large values in a truncated exponential fashion. This allows us to have a very
complex output signal. In our case the value of $k$ is 10000.

The voltage $V(t)$ across the junction is not taken constant. This voltage
will be produced by a chaotic system. In our case we selected the Chua's
circuit \cite{Matsumoto}. For this, we have implemented the Chua's circuit
following the recipe of Ref. \cite{Kennedy}.

\begin{figure}[ptb]
%%\epsfysize=12.0truecm
%%\epsfxsize=12.0truecm
%%\epsffile{fig9.eps}
\centerline{\includegraphics[width=8.0cm,angle=-90]{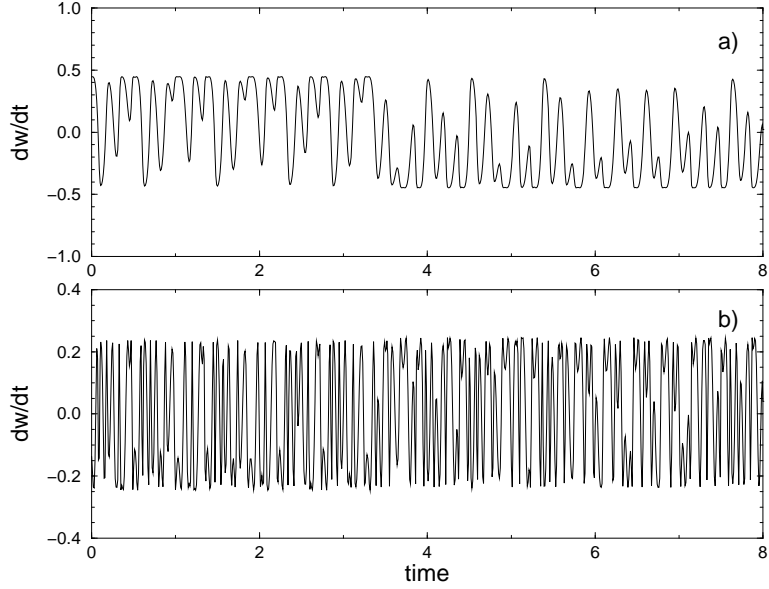}}  \vspace
{5mm}\caption{Time-series generated by the dynamical system 
(\ref{exp32})-(\ref{exp35}): (a) $g\left(  x\right)  =4X^{3}-3X$;
(b) $g\left(  x\right)  =\frac{1}{256}\left( 88179X^{11}-230945X^{9}
+218790X^{7}-90090X^{5}+15015X^{3}-693X\right)$.}%
\label{Fig.9}%
\end{figure}

The scheme of the Chua's circuit constructed by us can be found in Fig. 1 of
Ref. \cite{Kennedy}.

The following components were used: $C_{1}=10$ nF, $C_{2}=100$ nF, $L=19$ mH,
and $R$ is a $2.0$ K$\Omega$ trimpot.

Chua's diode was built using a two-operational-amplifier configuration
suggested in \cite{Kennedy}.

In our experiment, the voltage in $C_{1}$ was used as the driving signal for
the Josephson junction. We were interested in the famous double scroll
attractor attained with $R\approx1880$ $\Omega$.

The differential equations that describe our experimental system are the following:%

\begin{equation}
\frac{dV_{1}}{dt}=6.3\left(  V_{2}-V_{1}\right)  -9f(V_{1}),\label{exp36}%
\end{equation}%

\begin{equation}
\frac{dV_{2}}{dt}=0.7\left(  V_{1}-V_{2}\right)  +I_{L},\label{exp37}%
\end{equation}%

\begin{equation}
\frac{dI_{L}}{dt}=-7V_{2},\label{exp38}%
\end{equation}%

\begin{equation}
\frac{d\phi}{dt}=kV_{1},\label{exp39}%
\end{equation}%

\begin{equation}
\frac{dQ}{dt}=\sin\phi,\label{exp40}%
\end{equation}
where $f(V_{1})=-0.5\left[  V_{1}+0.3\left(  \left|  V_{1}+1\right|  -\left|
V_{1}-1\right|  \right)  \right]  $ and $k=10^{4}$. Notice that this system 
has been rewritten using adimensional variables (see Ref. \cite{Kennedy}).

The results of the experiments are shown in Fig. 10 which is the first-return
map of the time-series data produced by direct measurements of the junction
current. The time-intervals between measurements was 10 ms. This system can
produce unpredictable dynamics. Figure 11 shows the results of the
simulation of dynamical system (\ref{exp36})-(\ref{exp40}).

\begin{figure}[ptb]
%%\epsfysize=12.0truecm
%%\epsfxsize=6.0truecm
%%\rotatebox{270}{\epsffile{fig10.eps}}
\centerline{\includegraphics[width=8.0cm,angle=0]{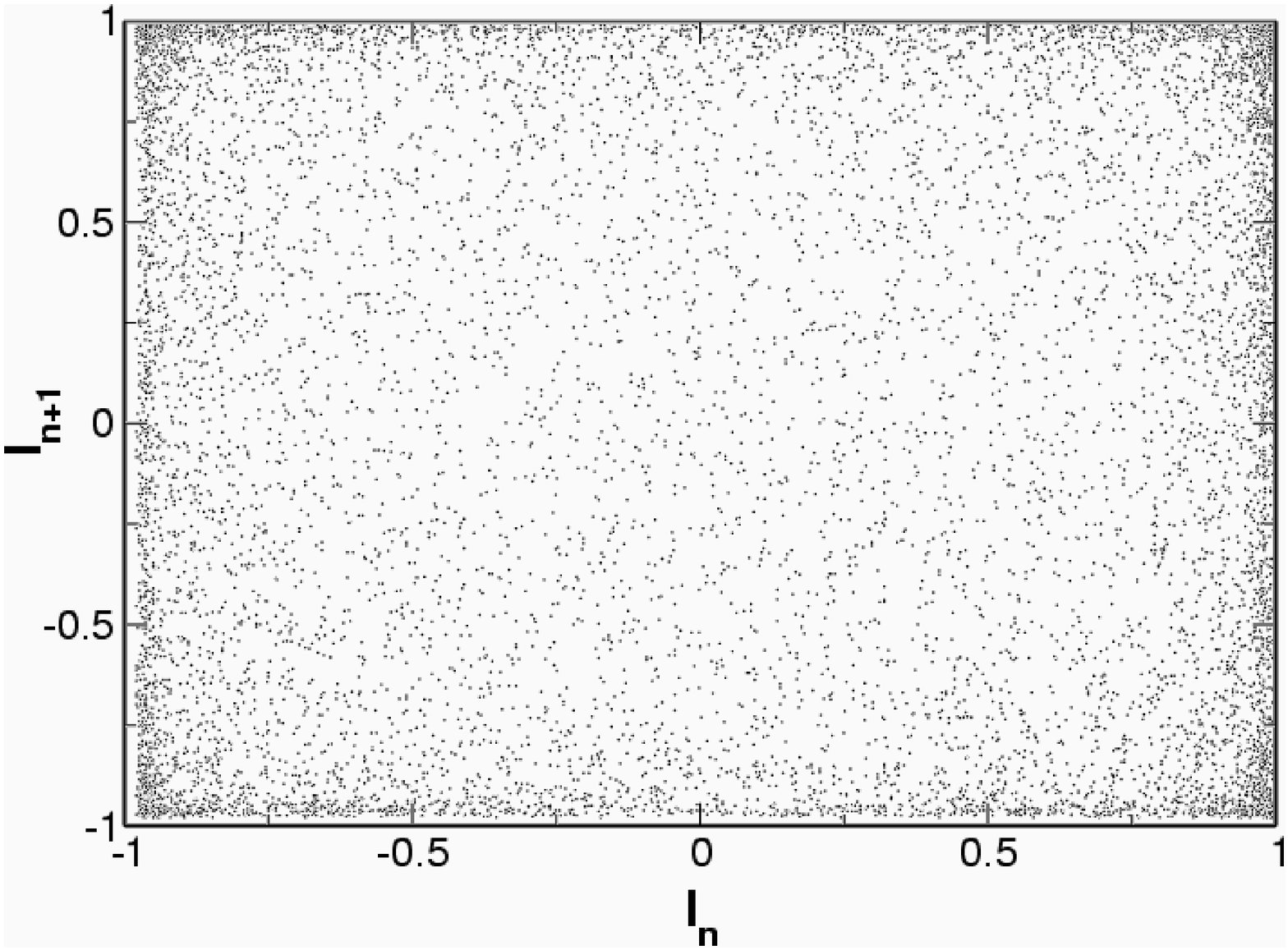}}  \vspace
{5mm}\caption{First-return map of the time-series generated with real data
from an experiment with an analog Josephson junction coupled to the Chua's
circuit.}%
\label{Fig.10}%
\end{figure}

\section{Conclusions}

In conclusion, we have shown that functions of type $X_{n}=P\left(  \theta
z^{n}\right)  $, where $P(t)$ is a periodic function and $z$ is a noninteger
number, can produce completely random numbers. Certain class of autonomous
dynamical systems can generate a similar dynamics. This dynamics presents
fundamental differences with the known chaotic systems. We have presented real
nonlinear systems that can produce this kind of random time-series. We have
reported the results of real experiments with nonlinear circuits containing
direct evidence supporting this phenomenon.

Besides the fundamental importance of these findings, these systems possess
many practical applications. For instance, game theory tells us that in
certain competitive situations the optimal strategy is a random behavior.
Specifically, it is necessary to limit the competition's ability to predict
our decisions. We can produce randomness using the discussed systems. Another
example is secure communications \cite{Pecora}. In this context, the most
important application of our systems is masking messages using random signals
\cite{Brown}. In some cases, when we use the usual chaotic systems, the
messages can be cracked because the time-series are not truly unpredictable.

Now we will analyze very general ideas.

Just to facilitate our discussion (because it is always important to have a
name), we will call the phenomenon studied in this paper ``deterministic
randomness''. The words ``deterministic randomness'' have been used
(metaphorically) in the past as a name for chaos. However, the known chaotic
systems are not random. So we think this is a good name for the present phenomenon.

Deterministic randomness imposes fundamental limits on prediction, but it also
suggests that there could exist causal relationships where none were
previously suspected.

Deterministic randomness demonstrates that a system can have the most
complicated behavior that emerges as a consequence of simple, nonlinear
interaction of only a few effective degrees of freedom.

\begin{figure}[ptb]
%%\epsfysize=12.0truecm
%%\epsfxsize=12.0truecm
%%\epsffile{fig11.eps}
\centerline{\includegraphics[width=8.0cm,angle=0]{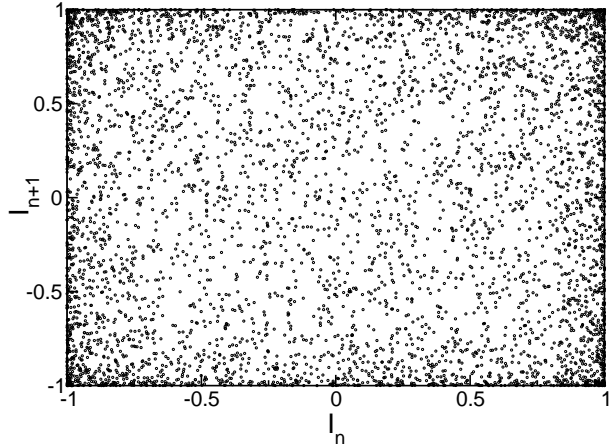}}  \vspace
{5mm}\caption{First-return map of the time-series generated by the dynamical 
system (\ref{exp36})-(\ref{exp40}). Here $I=\frac{dQ}{dt}$.}%
\label{Fig.11}%
\end{figure}

On one hand, deterministic randomness implies that if there is a phenomenon in
the World (whose mechanism from first principles is not known) described by a
dynamical system of type (\ref{exp8})-(\ref{exp10}) or
(\ref{exp11})-(\ref{exp13}), and the only observable is a physical
variable as $Z_{n}$, then the law of this phenomenon cannot be learnt from the
experimental data, or the observations. And, situations in which the
fundamental law should be inferred from the observations alone have not been
uncommon in physics.

On the other hand, the fact that this random dynamics is produced by a
relatively simple, well-defined autonomous dynamical system implies that many
random phenomena could be more predictable than have been thought.

Suppose there is a system thought to be completely random. From the
observation of some single variable, scientists cannot obtain the generation
law. However, suppose that in some cases, studying the deep connections of the
phenomenon, we can deduce a dynamical system of type (\ref{exp8})-(\ref{exp10})
or (\ref{exp11})-(\ref{exp13}). In these cases, some prediction is possible.

In any case, what is certain at this point is that some dynamical systems can
generate randomness on their own without the need for any external random input.


\begin{thebibliography}{9}                                                                                                %
\bibitem {Lorenz}E. N. Lorenz, J. Atmos. Sci. 20 (1963) 130.

\bibitem {Li}T. Y. Li and J. A. Yorke, Am. Math. Mon. 82 (1975) 985.

\bibitem {Feigenbaum}M. J. Feigenbaum, J. Stat. Phys. 19 (1978) 25.

\bibitem {May}R. M. May, Nature 261 (1976) 459.

\bibitem {Ford}J. Ford, Phys. Today 36 (1983) 40.

\bibitem {Crutchfield}J. P. Crutchfield, J. D. Farmer, N. H. Packard, and
R.\ S. Shaw, Sci. Am. 254 (1986) 46.

\bibitem {Grebogi}C. Grebogi, E. Ott, and J. A. Yorke, Science 238 (1987) 632.

\bibitem {Ott}E.Ott and M. Spano, Phys. Today 48 (1995) 34.

\bibitem {Takens}F. Takens, in Lecture Notes in Mathematics, 894, Ed. D. A.
Rand and L. S. Young, Springer-Verlag, Berlin, 1981.

\bibitem {Ruelle}D. Ruelle, Turbulence, strange attractors and chaos, World
Scientific, New Jersey, 1995

\bibitem {Farmer}J. D. Farmer and J. J. Sidorovich, Phys. Rev. Lett. 59 (1987)
845; G. Sugihara and R. M. May, Nature 344 (1990) 734; D. J. Wales, Nature 350
(1991) 485; A. A. Tsonis and J. B. Elsner, Nature 358 (1992) 217.

\bibitem {Gonzalez1}J. A. Gonz\'{a}lez and L. B. Carvalho, Mod. Phys. Lett. B
11 (1997) 521.

\bibitem {Nazareno}H. N. Nazareno, J. A. Gonz\'{a}lez, and I. Costa, Phys.
Rev. B 57 (1998) 13583.

\bibitem {Gonzalez2}J. A. Gonz\'{a}lez and R. Pino, Comp. Phys. Comun. 120
(1999) 109; J. A. Gonz\'{a}lez and R. Pino, Physica A 276 (2000) 425; J. A.
Gonz\'{a}lez, M. Mart\'{\i}n-Landrove, and L. Trujillo, Int. J. Bifurcation
and Chaos 10 (2000) 1867; J. A. Gonz\'{a}lez, L. I. Reyes,and L. E. Guerrero,
Chaos 11 (2001) 1.

\bibitem {Grassberger}P. Grassberger and I. Proccacia, Physica D 9 (1983) 189;
Wolf, J. B. Swift, H. L. Swinney, and J. A. Vastano, Physica D 16 (1985) 285.

\bibitem {Ferrenberg}A. M. Ferrenberg, D. P. Landau, Y. J. Wong, Phys. Rev.
Lett. 69 (1992) 3382.

\bibitem {James1}F. James, Comput. Phys. Commun. 60 (1990) 329.

\bibitem {James2}F. James, Chaos, Solitons and Fractals 6 (1995) 221.

\bibitem {Collins}J. J. Collins, M. Fanciulli, R. G. Hohlfeld, D. C. Finch, G.
v. H. Sandri, E. S. Shtatland, Comput. Phys. 6 (1992) 630.

\bibitem {Knuth}D. E. Knuth, Seminumerical Algorithms, The Art of Computer
Programming, Vol. 2, Addison-Wesley, Reading, MA, 1981.

\bibitem {Park}S. Park, K. Miller, Commun. ACM 31 (1988) 1192.

\bibitem {Doob}J. L. Doob, Stochastic Processes, Wiley, New York, 1991.

\bibitem {Mark}L. Mark Berliner, Statist. Sci. 7 (1992) 69.

\bibitem {D'Souza}R. M. D'Souza, Y. Bar-Yam, and M. Kardar, Phys. Rev. E 57
(1998) 5044.

\bibitem {Nogues}J. Nogu\'{e}s, J. L. Costa-Kr\"{a}mer, and K. V. Rao, Physica
A 250 (1998) 327.

\bibitem {Fisher}M. E. Fisher, Physica A 263 (1999) 554.

\bibitem {Marsaglia1}G. Marsaglia, Proc. Natl. Acad. Sci. 61 (1968) 25.

\bibitem {Marsaglia2}G. Marsaglia, in Computer Science and Statistics: The
Interface, Ed. L. Billard, Elsevier, Amsterdam, 1985.

\bibitem {Marsaglia3}G. Marsaglia, B. Narasimhan, and A. Zaman, Comput. Phys.
Commun. 60 (1990) 345.

\bibitem {Vattulainen}I. Vattulainen, T. Ala-Nissila, and K. Kankaala, Phys.
Rev. Lett. 73 (1994) 2513.

\bibitem {Shchur}L. N. Shchur, J. R. Heringa, and H. W. J. Bl\"{o}te, Physica
A 241 (1997) 579.

\bibitem {Li2}T. Y. Li and J. A. Yorke, Nonlinear Anal. Theory Methods Appl. 2
(1978) 473.

\bibitem {Phatak}S. C. Phatak and S. S. Rao, Phys. Rev. E 51 (1995) 3670.

\bibitem {Kautz}R. L. Kautz, J. Appl. Phys. 86 (1999) 5794.

\bibitem {Proykova}A. Proykova, Comput. Phys. Commun. 124 (2000) 125.

\bibitem {Stojanovski1}T. Stojanovski and L. Kocarev, IEEE Trans. Circuits
Syst. I 48 (2001) 281.

\bibitem {Stojanovski2}T. Stojanovski and L. Kocarev, IEEE Trans. Circuits
Syst. I 48 (2001) 382.

\bibitem {Bernardini}R. Bernardini and G. Cortelazzo, IEEE Trans. Circuits
Syst. I (2001) 552.

\bibitem {Chua}L. O. Chua, C. A. Desoer, and E. S. Kuh, Linear and Nonlinear
Circuits, McGraw-Hill, New York, 1987.

\bibitem {Matsumoto}T. Matsumoto, L. O. Chua, and M. Komoro, IEEE Trans.
Circuits Syst. CAS-32 (1985) 797; Physica D 24 (1987) 97.

\bibitem {Rodriguez}A. Rodr\'{\i}guez-V\'{a}squez, J. L. Huertas, A. Rueda, B.
P\'{e}rez-Verd\'{u}, and L. O. Chua, Proc. IEEE 75 (1987) 1090.

\bibitem {Huang}A. S. Huang, L. Pirka, and M. Franz, Int. J. Bifurcation and
Chaos 6 (1996) 2175.

\bibitem {Barone}A. Barone and G. Paterno, Physics and Applications of the
Josephson Effect, Wiley-Interscience, New York, 1982; K. K. Likharev, Dynamics
of Josephson Junctions and Circuits, Gordon and Breach Science Publishers, New
York, 1986.

\bibitem {Kennedy}M. P. Kennedy, Frequenz 46 (1992) 66.

\bibitem {Werthamer}N. R. Werthamer and S. Shapiro, Phys. Rev. 164 (1967) 523.

\bibitem {Hamilton}C. A. Hamilton, Rev. Sci. Instrum. 43 (1972) 445.

\bibitem {Bak}C. K. Bak and N. F. Pedersen, Appl. Phys. Lett. 22 (1973) 149.

\bibitem {Magerlein}J. H. Magerlein, Rev. Sci. Instrum. 49 (1978) 486.

\bibitem {Pecora}L. M. Pecora and T. L. Carroll, Phys. Rev. Lett. 64 (1990)
821; T. L. Carroll and L. M. Pecora, Physica D 67 (1993) 126; K. Cuomo and A.
V. Oppenheim, Phys. Rev. Lett. 71 (1993) 65.

\bibitem {Brown}R. Brown and L. O. Chua, Int. J. Bifurcation and Chaos 6
(1996) 219; Y. Y. Chen, Europhys. Lett. 34 (1996) 24.
\end{thebibliography}
\end{document}